\documentclass[12pt,a4paper]{article}
\usepackage{amsmath}
\usepackage[dvips]{graphicx}
\input epsf
\usepackage{times}
\begin{document}
\begin{titlepage}
\title{Mass dependence in  vector--meson electroproduction}
\author{M. A. Perfilov$^a$, S. M. Troshin$^b$ \\[1ex]
\small \it ${(a)}$ Physics Department,\\
\small \it Moscow State University,\\
\small \it Vorobiovy Gory , Moscow,  119 899 Russia\\
\small  \it ${(b)}$ Institute for High Energy Physics,\\
\small  \it Protvino, Moscow Region, 142280, Russia}
\normalsize
\date{}
\maketitle

\begin{abstract}
We demonstrate that the explicit mass
dependence of the exponent  in the power--like energy behavior of the
vector--meson production cross-section in the processes of virtual photon interactions
with a proton $\gamma^*p\to Vp$ obtained in the off--shell extension of the
approach based on unitarity is in a quantitative
agreement with the high--energy HERA experimental data.
\end{abstract}
\end{titlepage}
\setcounter{page}{2}

\section*{Introduction}

Besides  the most known studies of DIS at low $x$
 important measurements of cross--sections
of the elastic
vector meson production were performed in the experiments H1 and
ZEUS at HERA \cite{zosa,melld}. As it follows from these data  the
 integral  cross section
of the elastic vector meson production increases with energy in a way
similar to the
$\sigma^{tot}_{\gamma^*p}(W^2, Q^2)$ dependence
 on  $W^2$ \cite{her}.
It appeared also that the growth of  the vector--meson
electroproduction cross--section  with energy is  steeper for
heavy vector mesons and when the virtuality
$Q^2$ increases.

In this note we show that the approach based on the
off-shell extension of the $s$--channel unitarity (cf. \cite{epj02} and references
therein) and its application
to the elastic vector meson production in the processes
$\gamma^*p\to Vp$ allows in particular to consider mass dependence of these processes.
 It appears that
the obtained mass and $Q^2$ dependencies are in a  quantitative
 agreement with the high--energy
HERA experimental data.

\section{Vector--meson electroproduction}
The extension of the $U$--matrix unitarization for the off-shell
scattering was considered in \cite{epj02}. It was supposed that
the virtual  photon fluctuates into a quark--antiquark
pair $q \bar q$ and this pair can
be treated as an effective virtual  vector meson state
in the processes with small Bjorken  $x$.
There were considered limitations the unitarity provides for the $\gamma^* p$--total
 cross-sections and geometrical effects in the
 energy dependence of $\sigma^{tot}_{\gamma^* p}$.
In particular, it was shown that an assumption of
 the $Q^2$--dependent constituent quark
 interaction radius    leads to the following
asymptotical dependence:
$\sigma^{tot}_{\gamma^* p}\sim (W^2)^{\lambda(Q^2)}$,
where $\lambda(Q^2)$ will be saturated at large values of $Q^2$.
This result is valid  when the interaction radius of the virtual
constituent quark is rising with virtuality $Q^2$.
The form corresponding to the virtual constituent quark
interaction radius was chosen as following
\begin{equation}\label{rqvi}
r_{Q^*}=\xi(Q^2)/m_Q.
\end{equation}
Thus, the dependence on virtuality $Q^2$ comes through the dependence of the
intensity of the virtual constituent
 quark interaction $g(Q^2)$
and the $\xi(Q^2)$, which determines the quark
interaction radius (in the on-shell limit $g(Q^2)\to g$ and
$\xi(Q^2)\to\xi$).

The reason for the rising
 interaction radius of the virtual
constituent quark with virtuality $Q^2$
 might be of a dynamical nature and it could originate from
 the emission of the additional $q\bar q$--pairs in the
 nonperturbative  structure of a constituent quark.
In this
approach constituent quark consists of a current quark
and the  cloud of quark--antiquark pairs of the different
flavors \cite{csn}.
Available experimental data
 are consistent with the $\ln Q^2$--dependence of the radius of this
 cloud.
The introduction of the $Q^2$ dependence into the interaction radius of a constituent
quark which in this approach consists of a current quark
and the  cloud of quark--antiquark pairs of the different
flavors is the main issue of the off--shell extension of the
model, which provides at large values of
$W^2$
\begin{equation}\label{totv}
\sigma^{tot}_{\gamma^* p}(W^2,Q^2)\propto G(Q^2)\left(\frac{W^2}{m_Q^2}
\right)^{\lambda (Q^2)}
\ln \frac{W^2}{m_Q^2},
\end{equation}
where
\begin{equation}\label{lamb}
\lambda(Q^2)=\frac{\xi(Q^2)-\xi}{\xi(Q^2)}.
\end{equation}
 The value and $Q^2$ dependence of the
 exponent $\lambda(Q^2)$ is related to the interaction radius
 of the virtual constituent quark. The value of parameter $\xi$
 in the model is determined by the slope of the differential cross--section
of elastic scattering at large $t$ \cite{lang}
and from the $pp$-experimental data it follows that $\xi=2$.
From the data for $\lambda(Q^2)$ obtained
at HERA  the ``experimental'' $Q^2$--dependence
 of the function
$\xi(Q^2)$  has been calculated \cite{epj02}:
\begin{equation}\label{ksiq}
\xi(Q^2)=\frac{\xi}{1-\lambda(Q^2)}.
\end{equation}
 The rise of the function $\xi(Q^2)$  is slow and consistent with
 $\ln Q^2$ extrapolation:
\[
\xi(Q^2)=\xi + a\ln\left(1+ \frac{Q^2}{Q_0^2}\right),
\]
where $a=0.172$ and $Q_0^2=0.265$ GeV$^2$.

The inclusion of heavy vector meson production into this
scheme is straightforward: the virtual photon fluctuates before
the interaction with proton into the heavy quark--antiquark pair
 which constitutes
the virtual heavy vector meson state. After an interaction with a proton
this state turns out into a real  heavy vector meson.

Integral exclusive (elastic) cross--section of vector meson production in
the process $\gamma^*p\to Vp$ when the vector meson in the final
state contains not necessarily  light quarks can be calculated directly:
\begin{equation}\label{elvec}
\sigma^{V}_{\gamma^* p}(W^2,Q^2)\propto G_{V}(Q^2)\left(\frac{W^2}
{{m_Q}^2}
\right)^{\lambda_{V} (Q^2)}
\ln \frac{W^2}{{m_Q}^2},
\end{equation}
where
\begin{equation}\label{lavm}
\lambda_{V}(Q^2)= \lambda (Q^2)\frac{\tilde{m}_Q}{\langle m_Q \rangle}.
\end{equation}
In Eq. (\ref{lavm}) $\tilde{m}_Q$ denotes the mass of the constituent
quarks from
the vector meson and $\langle m_Q \rangle$ is the mean constituent
quark mass
of  system of the vector meson and proton system.
For the on--shell scattering we have a familiar Froissart--like
asymptotic energy dependence
\begin{equation}\label{elvecon}
\sigma^{V}_{\gamma^* p}(W^2,Q^2)\propto \frac{\xi^2}
{{m_Q}^2}
\ln^2 \frac{W^2}{{m_Q}^2}.
\end{equation}
It is evident from Eq. (\ref{elvec}) that
$\lambda_{V}(Q^2)=\lambda(Q^2)$
for the light vector mesons.
In the case when the vector meson
is very heavy, i.e. $\tilde m_Q\gg m_Q$ we
have
\[
\lambda_{V}(Q^2)=\frac{5}{2}\lambda(Q^2).
\]
We conclude that the respective cross--section
rises faster than the corresponding cross--section
of the light vector meson production, e.g. Eq. (\ref{lavm}) results in
\[
\lambda_{J/\Psi}(Q^2)\simeq 2\lambda(Q^2).
\]

To perform a fit to the high--energy HERA experimental
data \cite{zosa,melld} we have chosen
the functional dependence of $G_{V}(Q^2)$ in the form
\begin{equation}\label{gv}
G_{V}(Q^2)=g\left(1+\frac{Q^2}{Q_0^2}\right)^{-a}.
\end{equation}
The agreement  of Eqs. (\ref{elvec}) and (\ref{elvecon}) with experiment
when the function $G_{V}(Q^2)$
has the form of Eq. (\ref{gv}) is illustrated by the Figs. (1-4). The values of the three
 adjustable parameters
$g$, $Q_0^2$ and $a$ are given in the Table 1.
\section*{Conclusion}
A quantitative agreement with the high--energy HERA experimental data on elastic
vector--meson electroproduction is in favor of relation (\ref{lavm}) which provides explicit
mass dependence of the exponent  in the power--like energy dependence of cross--sections.
It means that the dependence of the constituent quark
interaction radius in the form $r_{Q^*}=\xi(Q^2)/m_Q$ on its mass and virtuality
has an experimental support and corresponding
 non--universal energy dependence predicted  in \cite{epj02}
 does not  contradict to
the high--energy experimental data on elastic
vector--meson electroproduction.

\section*{Acnowledgement}
We are grateful to N. E. Tyurin for the useful comments and discussions
 and A.~Prokudin for
 communications on experimental data analysis.

\newpage
\begin{table}
\begin{center}
\begin{tabular}{||c||c||c||c||}\hline\hline
 {Meson} & {$g$, $\mu$b} & {$Q_0^2$} & {$a$} \\
\hline\hline
 $\rho$      & $1.22\cdot 10^{-3}$       & 0.66  & 2.84  \\ \hline
 $\omega$      & $1.20\cdot 10^{-4}$       & 0.71  & 2.52        \\ \hline
 $\phi$      & $1.11\cdot 10^{-4}$   & 0.76 & 2.87     \\ \hline
 $J/\psi$      & $7.87\cdot 10^{-6}$   & 0.86 & 1.87     \\ \hline \hline
\end{tabular}
\end{center}
\caption{The values of the adjustable parameters}
\end{table}

\begin{figure}[htb]
\begin{center}
\epsfxsize=4in \epsfysize=3in
 \epsffile{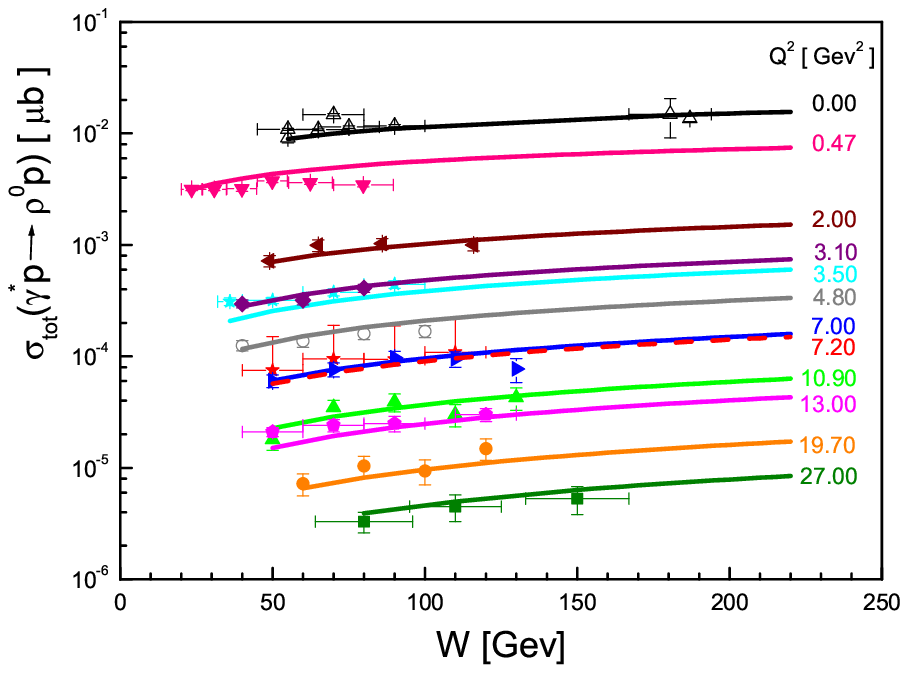}
\end{center}
 \caption[ro]{Energy dependence of the elastic
 cross--section of exclusive $\rho$--meson production.}
\label{fig:1}
\end{figure}
\begin{figure}[htb]
\begin{center}
\epsfxsize=4in \epsfysize=3in
 \epsffile{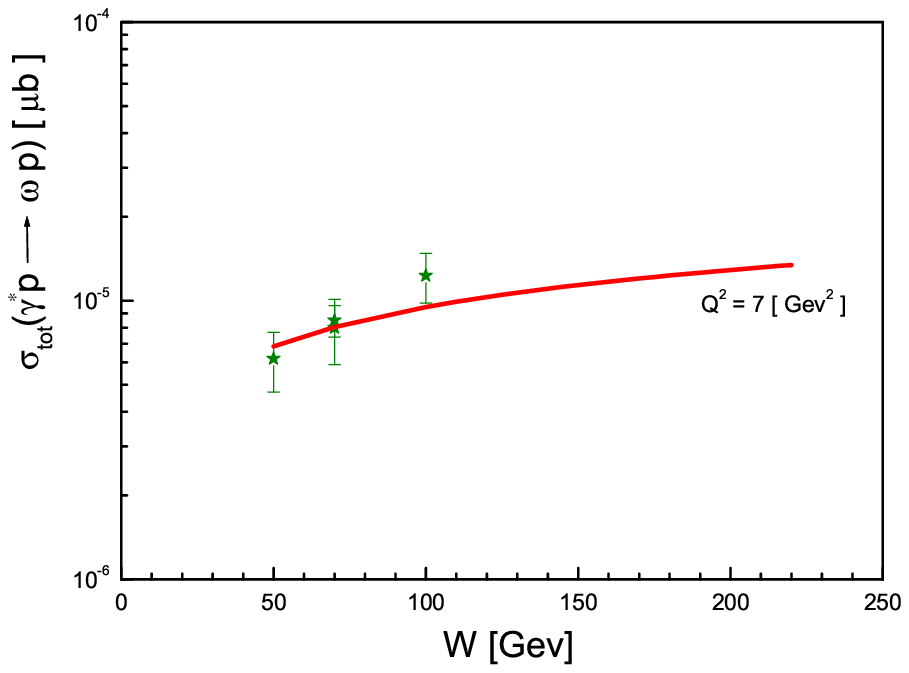}
\end{center}
 \caption[omega]{Energy dependence of the elastic
 cross--section of exclusive $\omega$--meson production.}
\label{fig:2}
\end{figure}
\begin{figure}[htb]
\begin{center}
\epsfxsize=4in \epsfysize=3in
 \epsffile{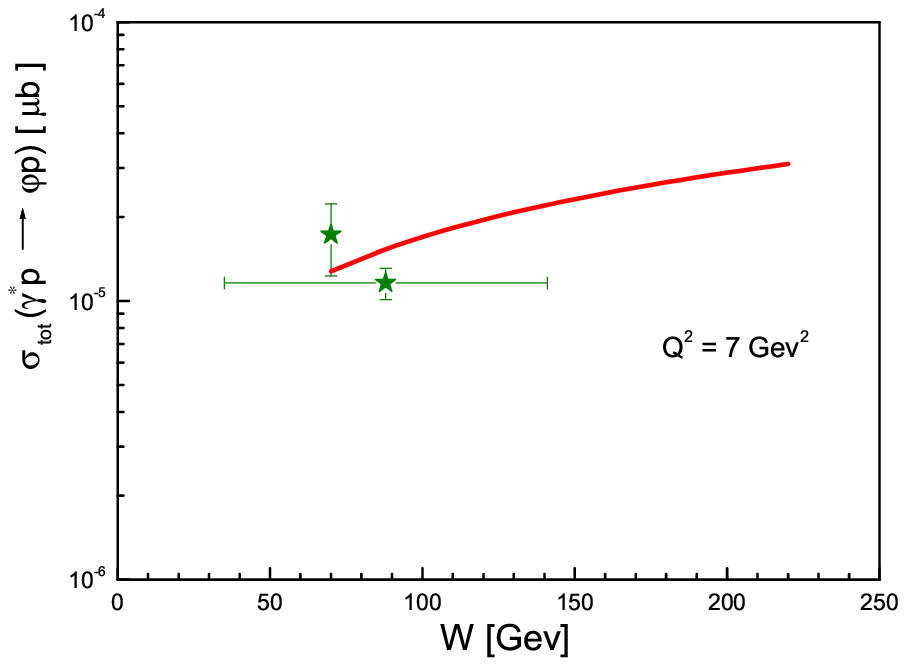}
\end{center}
 \caption[phi]{Energy dependence of the elastic
 cross--section of exclusive $\phi$--meson production.}
\label{fig:3}
\end{figure}
\begin{figure}[htb]
\begin{center}
\epsfxsize=4in \epsfysize=3in
 \epsffile{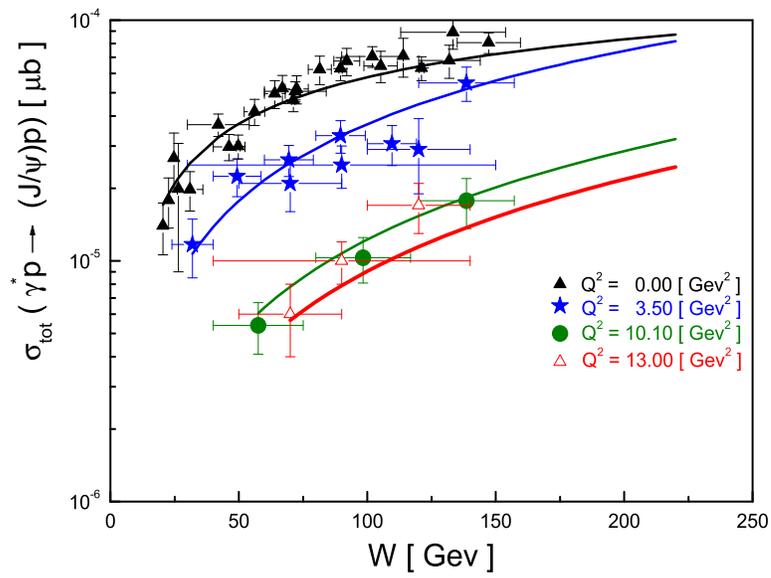}
\end{center}
 \caption[psi]{Energy dependence of the elastic
 cross--section of exclusive $J/\psi$ production.}
\label{fig:4}
\end{figure}

\end{document}